\documentclass[aps,twocolumn,showpacs,preprintnumbers,amsmath,amssymb,superscriptaddress,floatfix,nofootinbib]{revtex4}
\usepackage{lipsum}
\usepackage{graphicx}
\usepackage{epsfig}
\usepackage{epstopdf}
\usepackage{hyperref}
\usepackage{amsmath}
\usepackage{amsfonts}
\usepackage{amssymb}
\usepackage{caption}
\usepackage{subcaption}
\DeclareMathSizes{12}{12}{12}{12}

\begin{document}

\title{Predictions for $\Xi_b^- \to \pi^- (D_s^- ) \ \Xi_c^0 (2790) \left(\Xi_c^0 (2815) \right)$ and $\Xi_b^- \to \bar{\nu}_l l \ \Xi_c^0 (2790) \left(\Xi_c^0 (2815) \right)$}

\author{R. P. Pavao}
\email{rpavao@ific.uv.es}
\affiliation{Instituto de F\'\i sica Corpuscular (IFIC), Centro Mixto
CSIC-Universidad de Valencia, Institutos de Investigaci\'on de
Paterna, Apartado 22085, E-46071 Valencia, Spain}

\author{Wei-Hong Liang}
\email{liangwh@gxnu.edu.cn}
\affiliation{Department of Physics, Guangxi Normal University,
Guilin 541004, China}

\author{J. Nieves}
\email{jmnieves@ific.uv.es}
\affiliation{Instituto de F\'\i sica Corpuscular (IFIC), Centro Mixto
CSIC-Universidad de Valencia, Institutos de Investigaci\'on de
Paterna, Apartado 22085, E-46071 Valencia, Spain}

\author{E.~Oset}
\email{oset@ific.uv.es}
\affiliation{Departamento de
F\'{\i}sica Te\'orica and IFIC, Centro Mixto Universidad de
Valencia-CSIC Institutos de Investigaci\'on de Paterna, Aptdo.
22085, 46071 Valencia, Spain}

\date{\today}

\begin{abstract}
We have performed  calculations for the nonleptonic $\Xi_b^- \rightarrow \pi^- \ \Xi_c^0 (2790) \left(J=\frac{1}{2}\right)$ and $\Xi_b^- \rightarrow \pi^- \ \Xi_c^0 (2815) \left(J=\frac{3}{2}\right)$ decays and the same reactions replacing the $\pi^-$ by a $D_s^-$. At the same time we have also evaluated the semileptonic rates for $\Xi_b^- \rightarrow \bar{\nu}_l l \ \Xi_c^0 (2790)$ and $\Xi_b^- \rightarrow \bar{\nu}_l l \ \Xi_c^0 (2815)$.
We look at the reactions from the perspective that the $\Xi_c^0 (2790)$ and $\Xi_c^0 (2815)$ resonances are dynamically generated from the pseudoscalar-baryon and vector-baryon interactions. We evaluate ratios of the rates of these reactions and make predictions that can be tested in future experiments. We also find that the results are rather sensitive to the coupling of the $\Xi_c^*$ resonances to the $D^* \Sigma$ and $D^* \Lambda$ components.
\end{abstract}

\maketitle

\section{Introduction}
The introduction of chiral dynamics in the study of meson-baryon interactions \cite{ecker, ulfg} has allowed a rapid development in this field. A qualitative step forward was given by introducing unitarity in coupled channels, using the chiral Lagrangians as a source of the interaction \cite{weise,angels,ollerulf,carmen,hyodo}. In many cases the interaction is strong enough to generate bound states in some channels, which decay into the open states considered in the coupled channel formalism. The most renowned case is the one of the two $\Lambda(1405)$ states \cite{ollerulf,jido,ulfg2,carmen}.  The original works considered the interaction of pseudoscalar mesons with baryons, but the extension to vector mesons with baryons was soon done in Refs. \cite{angelsvec,sarkarvec}. The extension to vector mesons finds its natural framework in the use of the local hidden gauge Lagrangians \cite{hidden1,hidden2,hidden4}, which extend the chiral Lagrangians and accommodate vector mesons.

The mixing of pseudoscalar-baryon ($PB$) and vector-baryon ($VB$) channels in that framework was done in Ref. \cite{javier} in the light sector, and was extended to the charm sector in Refs. \cite{uchinoop,uchinohid}. An alternative approach to this mixing has been undertaken in Ref. \cite{romanets}, where the chiral Weinberg-Tomozawa (WT) meson-baryon interaction was extended to four flavors. Such an extension begins with the SU(8) spin-flavor symmetry group, including some symmetry breaking terms, and it  reduces to the SU(3) WT Hamiltonian when light pseudoscalar mesons are
involved, thus respecting chiral symmetry, while  heavy-quark spin symmetry (HQSS) is fulfilled in the heavy-quark sector.

One case where the relevance of the mixing is found is in the description of the $\Lambda_c(2595) (\frac{1}{2}^-)$ and $\Lambda_c(2625)(\frac{3}{2}^-)$. In early works on the subject, the  $\Lambda_c(2595)$ appeared basically as a $DN$ molecule \cite{lutz,mizutani}, but both in Refs. \cite{romanets} and \cite{uchinoop} a coupling to the $D^* N$ component was found with similar strength. On the other hand the $\Lambda_c(2625)$ appears from the  $\Sigma^*_c\pi-D^* N$ coupled-channel interaction in $S$-wave.

Support for the relevance of the vector-baryon components in these states was recently found in Refs. \cite{weihong,weisemi}. In Ref. \cite{weihong} the   $\Lambda_b \rightarrow \pi^- \Lambda_c(2595)$ and $\Lambda_b \rightarrow \pi^- \Lambda_c(2625)$ decays were studied and good agreement with experiment was found for the ratio of the two partial decay widths. The role of the $D^* N$ was found very important, to the point that if the sign of the coupling of the $D^* N$ to the $\Lambda_c(2595)$ was changed, the ratio of partial decay widths was in sheer disagreement with experiment. In Ref. \cite{weisemi} the semileptonic  ${\Lambda}_b \rightarrow \bar{\nu}_l l \Lambda _c(2595)$ and ${\Lambda}_b \rightarrow \bar{\nu}_l l \Lambda _c(2625)$ decay-modes were studied and the ratio of the partial decay widths was also found in agreement with experiment. Once again, reversing the sign of the $D^* N$ coupling to the $\Lambda_c(2595)$ led to results incompatible with experiment.

In the present work, we retake the ideas of Refs. \cite{weihong,weisemi} and apply them to the study of the  $\Xi_b^- \rightarrow \pi^- \ \Xi_c^0 (2790) (\frac{1}{2}^-)$, $\Xi_b^- \rightarrow \pi^- \ \Xi_c^0 (2815) (\frac{3}{2}^-)$, $\Xi_b^- \rightarrow D_s^- \Xi_c^0 (2790)$, $\Xi_b^- \rightarrow D_s^- \Xi_c^0 (2815)$, $\Xi_b^- \rightarrow \bar{\nu}_l l \ \Xi_c^0 (2790)$ and  $\Xi_b^- \rightarrow \bar{\nu}_l l \ \Xi_c^0 (2815)$ decays. The $\Xi_c^0 (2790) (\frac{1}{2}^-)$ and $\Xi_c^0 (2815) (\frac{3}{2}^-)$ play an analogous role to the $\Lambda_c(2595) (\frac{1}{2}^-)$ and $\Lambda_c(2625) (\frac{3}{2}^-)$, substituting the $u$-quark by an $s$-quark. In Ref. \cite{romanets} the couplings of the  $\Xi_c^0 (2790)$ and $\Xi_c^0 (2815)$ to the different coupled channels were evaluated for both pseudoscalar-baryon and vector-baryon components, in particular the $D \Lambda$, $D^* \Lambda$, $D \Sigma$, $D^* \Sigma$ which will be those needed in the decays mentioned above. We will adapt the formalism developed in Refs. \cite{weihong,weisemi} to the present case and will make predictions for these partial decay modes, which are not yet measured.

\section{Formalism}

\begin{figure}[h]
\centering
  \includegraphics[scale = 0.58]{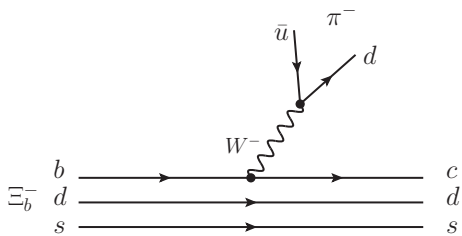}
  \captionsetup{justification=raggedright}
  \caption{Diagrammatic representation of the weak decay $\Xi_b^- \rightarrow \pi^- \Xi_c^*$.}
  \label{fig:decay1}
\end{figure}

We follow the steps of Ref. \cite{liang} for the weak decay of $B$ mesons leading to  hadronic resonances in the final state, generalized to the weak decay of $\Lambda_b$ baryons into baryonic resonances in Ref. \cite{mai}. In this latter study, the ${\Lambda}_b \rightarrow J/ \psi K^- p$ and ${\Lambda}_b \rightarrow J/ \psi \pi \Sigma$ reactions in the region of the $\Lambda(1405)$ resonance were studied, and predictions were made for the $K^- p$ invariant mass distribution, which were confirmed by experiment later in the LHCb work disclosing pentaquark states \cite{lhcb}. The analysis of Ref. \cite{mai} also predicted that the $K^- p$ and $\pi \Sigma$ would be produced with isospin $I=0$, which was also confirmed in Ref. \cite{lhcb} since their partial wave analysis only gave $J/ \psi$ and $\Lambda^*$ states. Work along the same lines as Ref. \cite{mai} was done in Ref. \cite{miyahara} in the decay of $\Lambda_c$ leading to $\Lambda(1405)$ and $\Lambda(1670)$, and in Ref. \cite{feijoo} in the   $\Lambda_b \rightarrow J/ \psi K \Xi$ reaction. The scheme of Ref. \cite{mai} applied to  the present case proceeds as depicted in Fig. \ref{fig:decay1}.

The first point to take into account is that in the $\Xi_b^-$ baryon, the $ds$ pair has spin $S=0$. Symmetry of the wave function requires the flavour combination $ds-sd$, and color provides the antisymmetry. The next step is the hadronization of the final $cds$ state into meson-baryon pairs.

\begin{figure}[h!]
\centering
  \includegraphics[scale = 0.58]{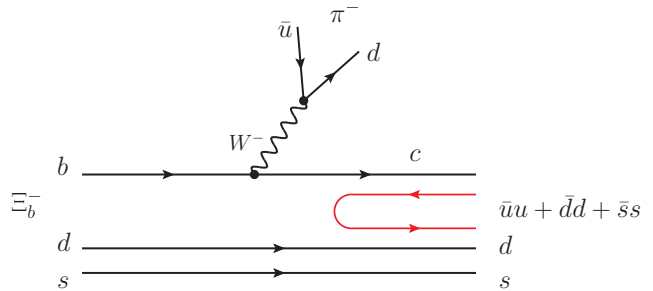}
  \caption{Hadronization after the weak process in Fig. \ref{fig:decay1} to produce a meson-baryon pair in the final state.}
  \label{fig:decay2}
\end{figure}

We must consider some basic facts:
\begin{enumerate}
\item The $ds$ quarks are spectators in the process. They have $S=0$ and come in the combination $\frac{1}{\sqrt{2}} \left(ds-sd \right)$.
\item  We will consider only final $\Xi_c^*$ resonances with negative parity, and generated from the meson-baryon interaction in $S$-wave. Since the pair $ds$ has positive parity, the $c$ quark  must carry
the negative parity and hence it will be produced in $P$-wave ($L=1$) in the weak interaction diagram depicted in Fig. \ref{fig:decay1}.
\item The $c$ quark will be incorporated into a final $D \left( D^* \right)$ meson and thus will go back to its ground state. Hence, the hadronization, introducing $\left(\bar{u}u+\bar{d}d+\bar{s}s \right)$ with the quantum numbers of the vacuum, must involve the $c$ quark.
\end{enumerate}
With these constraints, the hadronization proceeds as shown in Fig. \ref{fig:decay2}.
\vspace{0.8cm}

Technically the hadronization is implemented as follows: The $\Xi_b^-$ state has a flavour function
\begin{equation}
\left|\Xi_b^- \right\rangle  \equiv \frac{1}{\sqrt{2}} \left|b \left(ds - sd\right) \right\rangle ,
\end{equation}
and after the weak decay, the $b$ quark is substituted by a $c$ quark and we will have a state
\begin{equation}
\left|H\right\rangle = \frac{1}{\sqrt{2}}\left| c \left(ds - sd\right)\right\rangle.
\end{equation}
With the hadronization, we will have now
\begin{equation}
 \left|H'\right\rangle = \frac{1}{\sqrt{2}} \sum_{i=1}^3  \left|P_{4 i} \ q_i \left(ds - sd\right)\right\rangle ,
\end{equation}
where $P_{ij}$ are the $q \bar{q}$ matrix elements.

Next we write the $q \bar{q}$ matrix in terms of the physical mesons, $P \rightarrow \phi$, with $\phi$ given by
\begin{widetext}
\begin{equation}
\label{Eq.:mmatrix}
\phi \equiv \begin{pmatrix}
\frac{1}{\sqrt{2}} \pi^0 + \frac{1}{\sqrt{3}} \eta + \frac{1}{\sqrt{6}} \eta ' & \pi^+ &  K^+  & \bar{D}^0\\
\pi^- & -\frac{1}{\sqrt{2}} \pi^0 + \frac{1}{\sqrt{3}} \eta + \frac{1}{\sqrt{6}} \eta ' & K^0 & D^-\\
K^- & \bar{K}^0 & -\frac{1}{\sqrt{3}} \eta + \sqrt{\frac{2}{3}} \eta ' & D_s^-\\
D^0 & D^+ & D_s^+  & \eta_c
\end{pmatrix}.
\end{equation}
Then we can write
\begin{equation}
\label{Eq.:hadronstate}
\left| H' \right\rangle = \frac{1}{\sqrt{2}} \left[\left| D^0 u  \left(ds - sd\right)\right\rangle+ \left|D^+ d  \left(ds - sd\right)\right\rangle+\left|D_s^+ s  \left(ds - sd\right)\right\rangle \right].
\end{equation}
\end{widetext}
The last state in Eq. \eqref{Eq.:hadronstate} contains two extra $s$ quarks and corresponds to a more massive component that we omit in our study.

Next we see that we have a mixed antisymmetric component for the baryonic states of three quarks. If we evaluate the overlap with the mixed antisymmetric representations of the $\Sigma^-$,  $\Sigma^0$, $\Lambda^0$ states \cite{close}, we find
\begin{equation}
\left| H' \right\rangle = \frac{1}{\sqrt{2}}\left|D^0 \Sigma^0 \right\rangle+\left|D^+ \Sigma^- \right\rangle-\frac{1}{\sqrt{6}}\left|  D^0 \Lambda\right\rangle.
\end{equation}
Yet, we have to be careful here with the phase conventions. By looking at the phase convention of Ref. \cite{close} and the one inherent in the baryon octet matrix,
\begin{equation}
\label{Eq.:bmatrix}
B=\begin{pmatrix}
\frac{1}{\sqrt{2}} \Sigma^0+\frac{1}{\sqrt{6}} \Lambda & \Sigma^+ & p\\
\Sigma^- & -\frac{1}{\sqrt{2}} \Sigma^0+\frac{1}{\sqrt{6}} & n\\
\Xi^- & \Xi^0 & - \frac{2}{\sqrt{6}} \Lambda
\end{pmatrix},
\end{equation}
which is used in the chiral Lagrangians, one can see that one must change the phases of $\Sigma^+$, $\Lambda$, $\Xi^0$ from Ref.~\cite{close} to agree with the chiral Lagrangians\footnote{One way to see this is to take the singlet baryon state of Ref. \cite{close} with a minus sign, introduce the hadronization with $\bar{u}u+\bar{d}d+\bar{s}s$ as we have done before and see the meson-baryon content. The relative phases are deduced by comparing  this result with the SU(3) singlet $\text{Tr}\left(B \cdot \phi \right)$, obtained with the nonet of mesons in Eq. \eqref{Eq.:mmatrix} for $\phi$ (taking only the $3 \times 3$ part of the matrix), and Eq. \eqref{Eq.:bmatrix} for $B$. The matrix $\phi$ contains also a singlet of mesons, the octet matrix is the same putting in the diagonal $\left(\frac{\pi^0}{\sqrt{2}} + \frac{{\eta}_{8}}{\sqrt{6}}, -\frac{\pi^0}{\sqrt{2}}+\frac{\eta_8}{\sqrt{6}},-\frac{2 \eta_8}{\sqrt{6}} \right)$. Two alternative derivations are done in the Appendix of Ref.~\cite{Miyahara:2016yyh} with the same conclusions.}.

With this clarification about the phases, the state that we obtain consistent with the chiral convention is:
\begin{equation}
\left|H' \right \rangle=\frac{1}{\sqrt{2}}\left|D^0 \Sigma^0 \right\rangle+\left|D^+ \Sigma^- \right\rangle+\frac{1}{\sqrt{6}}\left|  D^0 \Lambda\right\rangle.
\end{equation}
We also mention the phase convention for mesons in terms of isospin states, where $\left|\pi^+ \right\rangle = - \left| 1,1 \right\rangle$,  $\left|K^- \right\rangle = - \left| \frac{1}{2},-\frac{1}{2} \right\rangle$, $\left|D^0 \right\rangle = - \left| \frac{1}{2},-\frac{1}{2} \right\rangle$, and for baryons $\Sigma^+=- \left| 1,1 \right\rangle$, $\Xi^-=- \left| \frac{1}{2},-\frac{1}{2} \right\rangle$.

In terms of isospin, $\left|H' \right\rangle$ can be written as
\begin{equation}
\label{Eq.:isoH}
\left|H' \right\rangle = -\sqrt{\frac{3}{2}} \left| \Sigma D (J=\frac{1}{2}) \right\rangle+\frac{1}{\sqrt{6}} \left| \Lambda D (J=\frac{1}{2}) \right\rangle.
\end{equation}
For $D^*$ production the flavour counting is the same and we would have the same combination substituting $D$ by $D^*$.

\subsection{The weak vertex}
One must evaluate the weak transition matrix elements. For this we follow the approach in Ref. \cite{weihong}. The vertex $W^- \rightarrow \pi^-$ is of the type \cite{gasser,scherer}
\begin{equation}\label{eq:LWpi}
\mathcal{L}_{W \pi} \sim W^{\mu} \partial_{\mu} \phi,
\end{equation}
while the $bcW$ vertex is of the type
\begin{equation}\label{eq:Lqwq}
\mathcal{L}_{q W q} \propto \bar{q}_{\text{fin}} W_{\mu} \gamma^{\mu}(1-\gamma_5) q_{\text{in}}.
\end{equation}
Since we are dealing with heavy quarks, as in Ref.~\cite{weihong} we keep the dominant terms in a nonrelativistic expansion:  $\gamma^0$ and $\gamma^i \gamma^5 \ \left(i=1,2,3\right)$. Thus,  combining the two former vertices  we obtain a structure for the weak transition at the quark level of the type
\begin{equation}
\label{Eq.:weaktrans}
V_P \sim q^0 + \vec{\sigma} \cdot \vec{q},
\end{equation}
with $q^{\mu}$ the four-momentum of the pion.

In Ref. \cite{weihong} the operator in Eq. \eqref{Eq.:weaktrans}, which acts at the quark level between the $b$ and $c$ quarks, was converted into an operator acting over the $\Lambda^*_c$ and $\Lambda_b$ at the macroscopical level with the result
\begin{equation}
\label{Eq:13}
V_P \sim \left\{ \left(i \frac{q^0}{q} \vec{\sigma} \cdot \vec{q} + i q \right) \delta_{J,\frac{1}{2}}-i \frac{q^0}{q} \sqrt{3} \ \vec{S}^+ \cdot \vec{q} \ \delta_{J,\frac{3}{2}} \right\} \text{ME}(q),
\end{equation}
where $\vec{S}^+$ is the spin transition operator from spin $\frac{1}{2}$ to spin $\frac{3}{2}$ normalized such that
\begin{equation}
\left < M' \right| S^+_{\mu} \left| M \right> = \mathcal{C}(\frac{1}{2}, \, 1, \, \frac{3}{2};\, M, \, \mu, \, M'),
\end{equation}
with $\mu$ in the spherical basis and $\mathcal{C}(\frac{1}{2}, \, 1, \, \frac{3}{2};\, M, \, \mu, \, M')$ the Clebsch-Gordan coefficients. In addition, ME$(q)$ is the quark matrix element involving the radial wave functions (here we do the same as in Ref.~\cite{weihong}, but  the macroscopic states are $\Xi_c^*$ and $\Xi_b$ respectively),
\begin{equation}
\label{Eq.:matel}
\text{ME}(q)=\int {\rm d}r \ r^2 j_1(qr) \phi_{\text{in}}(r) \phi^*_{\text{fin}}(r),
\end{equation}
where $j_1(qr)$ is a spherical Bessel function and $\phi_{\text{in}}(r)$ is the radial wave function of the $b$ quark in $\Xi^-_b$ and $\phi_{\text{fin}}(r)$ the radial wave function of the $c$ quark, prior to the hadronization, which is in an excited $L=1$ state.

Since we require ratios of production rates, the matrix element ME$(q)$ cancels in the ratio and what matters to differentiate the cases with spin $\frac{1}{2}$ and $\frac{3}{2}$ is the operator in Eq. \eqref{Eq:13}. One should note that the presence of the factor $j_1(qr)$ in Eq. \eqref{Eq.:matel} is due to the fact that the $c$ quark is created with $L=1$ as we discussed previously.

In Sect.~\ref{sec:Relat_Effec}, we will improve on the nonrelativistic approximation of Eq.~\eqref{Eq.:weaktrans}, but we already advance that the ratios of rates only change at the level of 1\% with respect to this nonrelativistic approximation.

\subsection{The spin structure in the hadronization}

The next issue is to see how the hadronization affects the cases of $DB$ or $D^* B$ (with $B= \Sigma, \Lambda$) production in spin $J=\frac{1}{2}$ or $\frac{3}{2}$. For this we follow again the approach of Ref. \cite{weihong}. The calculation proceeds as follows:
\begin{enumerate}
\item The $\bar{q}q$ pair is created with $J^P=0^+$. Since the $\bar{q}$ has negative intrinsic parity we need $L=1$ in the quarks to restore the positive parity and this forces the $\bar{q}q$ pair to come with spin $S=1$ to give $J=0$. This is the essence of the $^3P_0$ model \cite{close,oliver}.
\item Since what we want is to elaborate on the spin dependence of the matrix elements, we assume a zero range interaction, as is also done in similar problems like the study of pairing in nuclei \cite{brown,preston}.
\item Since the $d, \ s$ quarks are spectators and carry $J=0$, the total angular momentum of the $\Xi_c^*$ is the same as the angular momentum of the $c$ quark after the weak production.
\item The angular momentum of the $c$ quark and the $\bar{q}q$ pair are recombined to give $L'=0$, since all quarks are in their ground state in the $D \Sigma$, $D^* \Sigma$, $D \Lambda$, and $D^* \Lambda$ final states. The total angular momentum of the $c$ quark and that of the $\bar{q}$ of the  $\bar{q}q$ pair are recombined to give $j=0,1$, for the $D$ or $D^*$ production. The total angular momentum of the $q$ from the $\bar{q}q$ pair determines the spin of the baryon $\Xi_c^*$ since the $ds$ quarks carry spin zero. The Clebsch-Gordan coefficients appearing in the different combinations are recombined to give a Racah coefficient \cite{rose} and the final result is (see Eq.~(24) of Ref.~\cite{weihong})
\begin{eqnarray}\label{Eq.:CGcoef}
   && \left|J M; c \right> \left|0 0; \bar{q}q \right>_{^3P_0} \left|0  0; ds \right>  \nonumber \\
  &=& \sum_j \mathcal{C}(j,J) \left|J, \, M; \text{meson-baryon}\right>,
\end{eqnarray}
where the coefficients $\mathcal{C}(j,J)$ are given in Table \ref{tab:tab1}.
\end{enumerate}
\begin{table}[tbp!]
\centering
\caption{$\mathcal{C}(j,J)$ coefficients in Eq. \eqref{Eq.:CGcoef}.\label{tab:tab1}}
\begin{tabular}{l|cc}
\hline\hline
$\mathcal{C}(j,J)$ & $J=\frac{1}{2}$ & $J=\frac{3}{2}$ \\ \hline
(pseudoscalar) $j=0~$ & $\frac{1}{4 \pi} \frac{1}{2}$ & 0 \\
(vector) $j=1$ & ~~$\frac{1}{4 \pi} \frac{1}{2\sqrt{3}}$~~ & $-\frac{1}{4 \pi} \frac{1}{\sqrt{3}}$ \\
\hline\hline
\end{tabular}
\end{table}

\begin{figure}[h!]
  \centering
  \includegraphics[scale = 0.58]{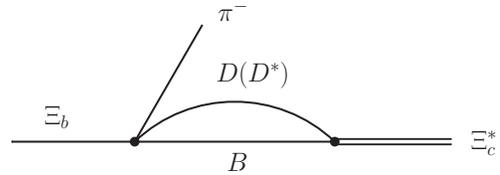}
  \captionsetup{justification=raggedright}
  \caption{Mechanism for the production of the $\Xi_c^*$ resonances by re-scattering of  $D \left(D^* \right) \Sigma \left(\Lambda\right)$ and coupling of the meson-baryon components to  $\Xi_c^*$.}
  \label{fig:decay3}
\end{figure}

What we have done so far is to obtain the angular structure of the mechanism for $DB \left(D^* B\right)$ production, but we finally want to have the production of the resonances $\Xi_c^0 (2790)$ and $\Xi_c^0 (2815)$. The way to produce these dynamically generated resonances is depicted in Fig. \ref{fig:decay3}.
It  involves the amplitudes for $\Xi_b \rightarrow \pi^- D \left(D^* \right)B$ production studied before, together with the $D \left(D^* \right)B$ loop functions and the couplings of the $\Xi_c^*$ resonance to these meson-baryon components.

The width for the $\Xi_b \rightarrow \pi^- \Xi^*_c$ decay is given by
\begin{equation}
\Gamma_{\Xi_b \rightarrow \pi^- \Xi_c^*} = \frac{1}{2 \pi}\, \frac{M_{\Xi_c^*}}{M_{\Xi_b}} \;q \; \overline{\sum} \sum |t|^2,
\end{equation}
with $q$ the momentum of the pion in the $\Xi_b$ rest frame.

By combining Eqs. \eqref{Eq.:isoH}, \eqref{Eq:13}, \eqref{Eq.:CGcoef}, we obtain
\begin{widetext}
\begin{eqnarray}\label{Eq.:sum12}
   J=\frac{1}{2}: \ \overline{\sum} \sum |t|^2 &=&  C^2 \left(q^2+\omega^2_{\pi} \right) \left|\frac{1}{2} \left(-\sqrt{\frac{3}{2}}\right) g_{R,\Sigma D}\,G_{\Sigma D}  + \frac{1}{2} \frac{1}{\sqrt{6}}  g_{R,\Lambda  D}\,G_{\Lambda D}\right. \nonumber \\
   && \left. + \frac{1}{2\sqrt{3}}\left(-\sqrt{\frac{3}{2}}\right) g_{R,\Sigma D^*}\,G_{\Sigma D^*} + \frac{1}{2\sqrt{3}}  \frac{1}{\sqrt{6}} g_{R,\Lambda  D^*}\, G_{\Lambda D^*}  \right|^2,
\end{eqnarray}
and
\begin{equation}
\label{Eq.:sum32}
J=\frac{3}{2}: \ \overline{\sum} \sum |t|^2 = C^2 \ 2\omega^2_{\pi} \left|\frac{1}{\sqrt{3}} \left(-\sqrt{\frac{3}{2}}\right)g_{R,\Sigma D^*}\, G_{\Sigma D^*} + \frac{1}{\sqrt{3}}  \frac{1}{\sqrt{6}} g_{R,\Lambda  D^*}\,G_{\Lambda D^*}  \right|^2,
\end{equation}
\end{widetext}
where $\omega_{\pi}$ is the pion energy $\sqrt{m_{\pi}^2+q^2}$, and $G_{BD}$, $G_{BD^*}$ are the loop functions for the propagator of $BD \left(B D^*\right)$ in the resonance formation mechanism of Fig. \ref{fig:decay3}, and $g_{R,BD \left(BD^*\right)}$ the coupling of the resonance $\Xi_c^*$ to any of the states $BD \left(BD^*\right)$. $C$ in Eqs. \eqref{Eq.:sum12} \eqref{Eq.:sum32} is a factor that contains the matrix element ME$(q)$ and constants of the weak interaction. Since the mass of the two $\Xi_c^*$ that we investigate are not very different, then we assume $C$ to be a constant that cancels in the ratio of the rates for the production of the two resonances. In this case we find
\begin{equation}
\label{Eq.:ratiorate}
R_1 \equiv \frac{\Gamma_{\Xi_b \rightarrow \pi^- \Xi_c (1)}}{\Gamma_{\Xi_b \rightarrow \pi^- \Xi_c (2)}} = \frac{M_{\Xi_c (1)} \;p_{\pi}(1) \;\overline{\sum}\sum |t|^2(1)}{M_{\Xi_c (2)} \;p_{\pi}(2)\; \overline{\sum}\sum |t|^2(2)},
\end{equation}
where $1,2$ refer to the $\Xi_c(2790)$ and $\Xi_c(2815)$ respectively.

The case of $D_s^-$ production instead of $\pi^-$ is identical. Instead of the $\bar{u}d$ coupling to the gauge boson $W$, we now have that of the $\bar{c}s$ pair, which is equally Cabbibo favoured and is proportional to $\cos \theta_C$ in both cases, with $\theta_C$ the Cabbibo angle. The only difference in this case is that the momentum of the $D_s^-$ is smaller than that in the case of pion production. The momenta of $D_s^-$ in the cases $\Xi_c (2790)$ and $\Xi_c (2815)$ are very similar and, hence, by analogy to Eq. \eqref{Eq.:ratiorate} we can write
\begin{equation}
\label{Eq.:ratiorate2}
R_2 \equiv \frac{\Gamma_{\Xi_b \rightarrow D_s^- \Xi_c (1)}}{\Gamma_{\Xi_b \rightarrow D_s^- \Xi_c (2)}} \!= \! \frac{M_{\Xi_c (1)} \,p_{D_s^-}(1) \,\overline{\sum}\sum |t|^2(1)}{M_{\Xi_c (2)} \,p_{D_s^-}(2) \,\overline{\sum}\sum |t|^2(2)},
\end{equation}
with $ p_{D_s^-}(1,2)$ evaluated for the $\Xi_c (2790)$ and $\Xi_c (2815)$ respectively, and  $\overline{\sum}\sum |t|^2(1,2)$ have to be reevaluated with the new momentum.

If we assume that ME$(q)$ is not very different in the case of $\pi^-$ or $D_s^-$ production we can also write
\begin{equation}
\label{Eq.:ratiorate3}
R_3 \equiv \frac{\Gamma_{\Xi_b \rightarrow D_s^- \Xi_c (1)}}{\Gamma_{\Xi_b \rightarrow \pi^- \Xi_c (1)}} = \frac{p_{D_s^-}(1)\; \overline{\sum}\sum |t|^2(1,D_s^-)}{p_{\pi^-}(1)\; \overline{\sum}\sum |t|^2(1,\pi^-)}.
\end{equation}
We expect this equation to hold only at the qualitative level since ME$(q)$ is not necessarily the same for these two different values of $q$.

\section{Semileptonic decay}
\begin{figure}[tbp!]
    \centering
    \begin{subfigure}[b]{0.42\textwidth}
          \includegraphics[width=\textwidth]{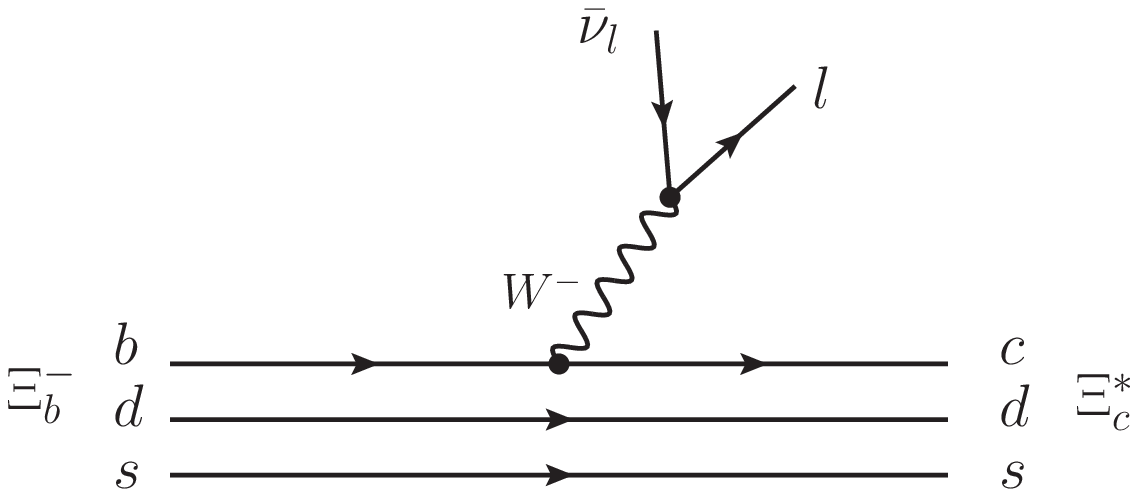}
        \caption{First step at quark level.}
        \label{fig:diag41}
    \end{subfigure}
    ~
    \begin{subfigure}[b]{0.48\textwidth}
          \includegraphics[width=\textwidth]{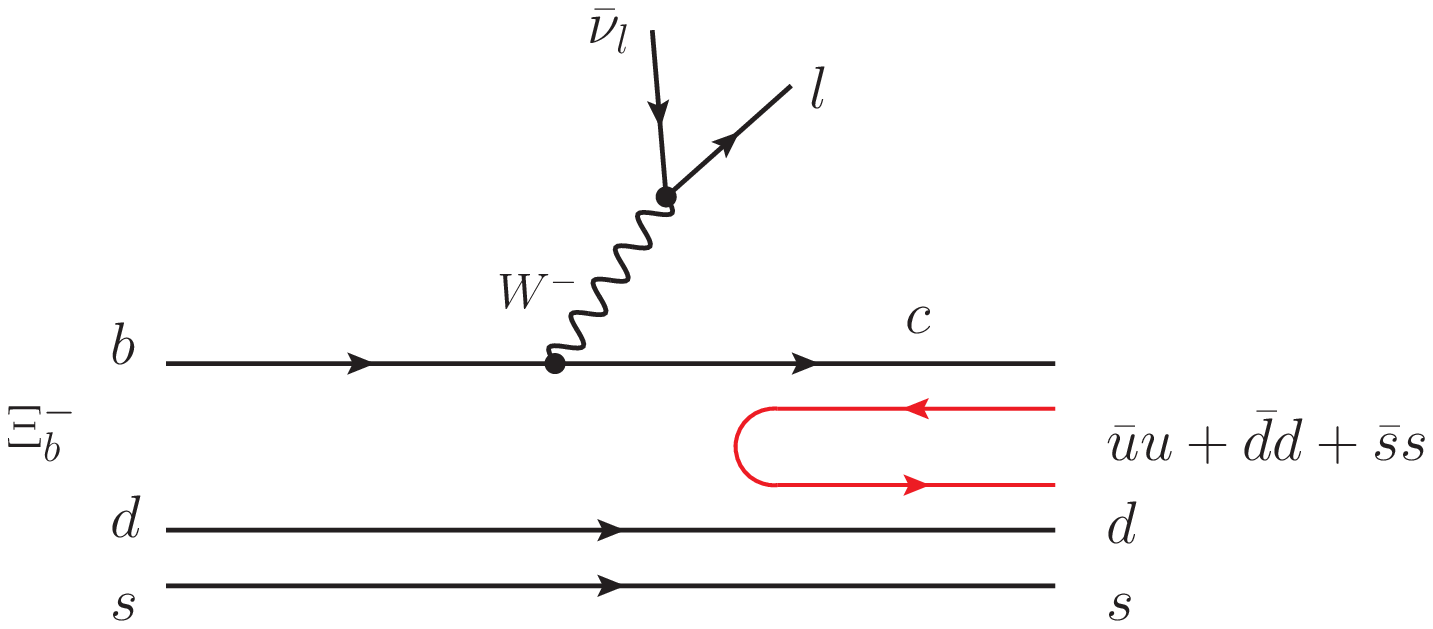}
        \caption{Hadronization to produce $D (D^*) B$.}
        \label{fig:diag42}
    \end{subfigure}
    ~
    \begin{subfigure}[b]{0.42\textwidth}
          \includegraphics[width=\textwidth]{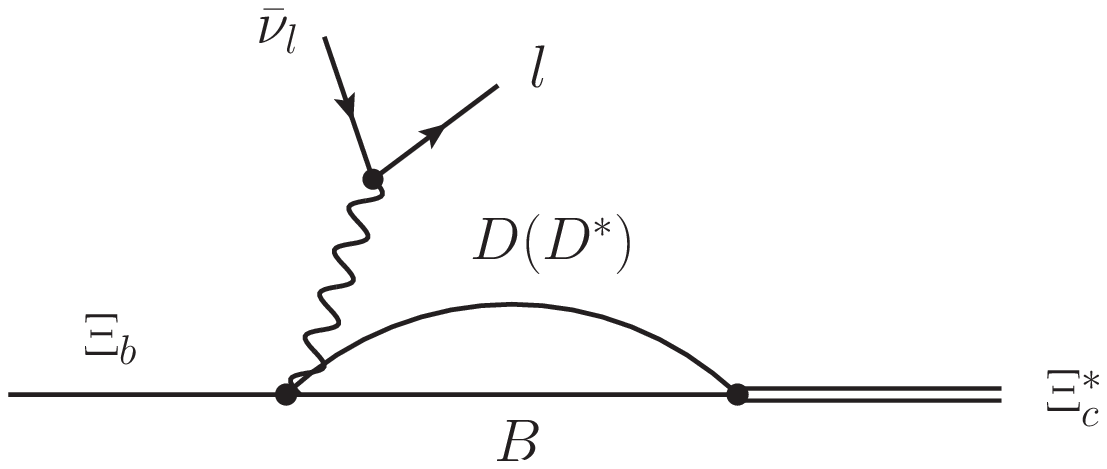}
        \captionsetup{justification=raggedright}
        \caption{Propagation of $D \left(D^*\right) B$ and coupling to the $\Xi_c^*$.}
        \label{fig:diag43}
    \end{subfigure}
    \captionsetup{justification=raggedright}
    \caption{Different steps of $\Xi_c^*$ production in the $\Xi_b \rightarrow \bar{\nu}_l l \Xi_c^*$ process.}\label{fig:animals}
\end{figure}

The semileptonic processes, $\Xi_b \rightarrow \bar{\nu}_l l \Xi_c^0(2790)$ and $\Xi_b \rightarrow \bar{\nu}_l l \Xi_c^0(2815)$ proceed in a similar way but instead of a $\pi^-$ we have $\bar{\nu}_l l$ production. The semileptonic decays of $B D$ hadrons along the lines described here have been studied in Refs. \cite{navarra,sekihara}. The weak decay of $\Lambda_c \rightarrow \bar{\nu}_l l \Lambda(1405)$ is addressed in Ref. \cite{ikeno} and the $\Lambda_b \rightarrow \bar{\nu}_l l \Lambda_c(2595)$ and $\Lambda_b \rightarrow \bar{\nu}_l l \Lambda_c(2625)$ in Ref. \cite{weisemi}. The first step for the $\Xi_b \rightarrow \bar{\nu}_l l \Xi_c^*$ reaction is shown in Fig. \ref{fig:diag41}

The only difference with the nonleptonic decay studied in the former sections is the coupling of $W$ to $\bar{\nu}_l l$. Following Ref. \cite{navarra} we have, for the combined $W \bar{\nu}_l l$ and $W c b$ vertices,
\begin{equation}
t' = -i G_F \;\frac{V_{bc}}{\sqrt{2}} \;L^{\alpha} Q_{\alpha},
\end{equation}
with $G_F$ the Fermi coupling constant, $V_{bc}$ the Cabbibo-Kobayashi-Maskawa matrix element for the $b \rightarrow c$ transition, and $L^{\alpha}, \ Q_{\alpha}$ the leptonic and quark currents:
\begin{subequations}
\begin{align}
&  L^{\alpha} = \bar{u}_l \gamma^{\alpha} (1-\gamma_5) u_{\nu_l}, \\
&  Q_{\alpha} = \bar{u}_c \gamma_{\alpha} (1-\gamma_5) u_{b}.
\end{align}
\end{subequations}

Once again we retain $\gamma^0$ and $\gamma^i \gamma_5$ from the quark matrix elements, which are the leading terms in a nonrelativistic reduction. Actually the $\bar{\nu}_l l$ pair comes out with a large momentum \cite{weisemi} and the momenta of the baryons are small.

The first step in Fig. \ref{fig:diag41} produces a different structure from Eq. \eqref{Eq.:weaktrans} in the nonleptonic case, and one finds (see Eqs.~(5),(6),(14) of Ref.~\cite{weisemi})
\begin{equation}\label{eq:LLQQ}
\sum_{\text{lepton pol.}} L^{\alpha} L^{\dagger \ \beta} Q_{\alpha} Q^{\dagger}_{\beta}=\frac{8}{m_{\nu} m_l} p_{\nu} p_l,
\end{equation}
where $p_{\nu}, p_l$ are the neutrino and lepton momenta in the $\Xi_b$ rest frame, and $m_\nu, m_l$ their masses. Note that we are using the field normalization of Mandl and Shaw \cite{mandl} and $\sum_\lambda u_\lambda (p) \bar u_\lambda (p)=(p\!\!\!\slash +m)/2m$. The masses $m_\nu, m_l$ in Eq.~\eqref{eq:LLQQ} get canceled in the formula of the width, Eq.~\eqref{Eq.:ratiomass}, and there are no problems even in the limit of small or zero neutrino mass.

\begin{figure}[btp!]
  \centering
  \includegraphics[scale = 0.24]{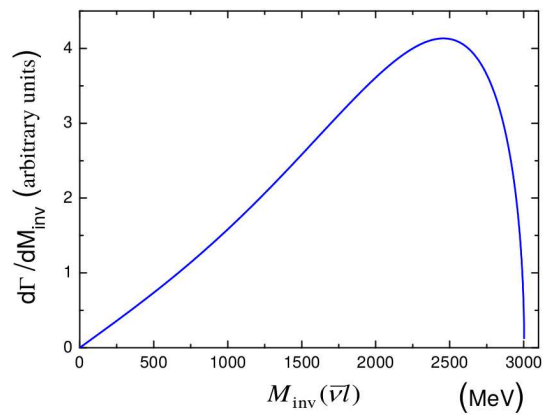}
  \captionsetup{justification=raggedright}
  \caption{The invariant mass distribution for $\bar{\nu}_l l$ in the $\Xi_b \rightarrow \bar{\nu}_l l \Xi_c (2790)$. The one for the $\Xi_b \rightarrow \bar{\nu}_l l \Xi_c (2815)$ decay is very similar.}
  \label{fig:ratiomass}
\end{figure}

The rest of the work needed is identical to the one in the nonleptonic case of the former sections. One can also do an angle integration analytically in the evaluation of $\Gamma$ and one finally obtains
\begin{equation}
\label{Eq.:ratiomass}
\frac{{\rm d} \Gamma}{{\rm d}M_{\text{inv}}(\bar{\nu}_l l)}= \frac{M_{\Xi_c^*}}{M_{\Xi_b}} 2 m_{\nu} 2 m_l \frac{1}{\left(2 \pi \right)^3} p_{\Xi_c^*} \tilde{p}_l \overline{\sum}\sum |t'|^2,
\end{equation}
where $p_{\Xi_c^*}$ is the $\Xi_c^*$ momentum in the $\Xi_b$ rest frame and $\tilde{p}_l$ the lepton momentum in the $\bar\nu_l l$ rest frame, and $\overline{\sum}\sum |t'|^2$ is given by \cite{weisemi}
\begin{widetext}
\begin{equation}
\overline{\sum}\sum |t'|^2 = C'^2 \frac{8}{m_{\nu} m_l} \frac{1}{M ^2_{\Xi_b}} \left(\frac{M_{\text{inv}}}{2}\right)^2 \left[\tilde{E}^2_{\Xi_b}-\frac{1}{3} \tilde{\vec{p}\,}^2_ {\Xi_b}\right] A_J V_{\text{had}}(J),
\end{equation}
with
\begin{equation}
\label{Eq.:sum122}
J=\frac{1}{2}: \ A_J V_{\text{had}}(J) = \left|\frac{1}{2} \left(-\sqrt{\frac{3}{2}}\right) g_{R,\Sigma D}\,G_{\Sigma D}  + \frac{1}{2} \frac{1}{\sqrt{6}} g_{R,\Lambda  D}\,G_{\Lambda D}
 + \frac{1}{2\sqrt{3}}\left(-\sqrt{\frac{3}{2}}\right) g_{R,\Sigma D^*}G_{\Sigma D^*} + \frac{1}{2\sqrt{3}}  \frac{1}{\sqrt{6}} g_{R,\Lambda  D^*}G_{\Lambda D^*}  \right|^2,
\end{equation}
and
\begin{equation}
\label{Eq.:sum322}
J=\frac{3}{2}:  \ A_J V_{\text{had}}(J)  = 2\left|\frac{1}{\sqrt{3}} (-\sqrt{\frac{3}{2}})g_{R,\Sigma D^*}\,G_{\Sigma D^*} + \frac{1}{\sqrt{3}}  \frac{1}{\sqrt{6}}g_{R,\Lambda  D^*}\, G_{\Lambda D^*}  \right|^2,
\end{equation}
\end{widetext}
where $G_{BD}, \ G_{BD^*}$ and $g_{R,BD}, \ g_{R, B D^*}$ are the same as in the nonleptonic decay and $C'$ is again a factor that contains the matrix element ME$(q)$ evaluated at the proper value of $q$. A novelty here is that $q$ is not constant when one integrates $\frac{{\rm d} \Gamma}{{\rm d}M_{\text{inv}}}$ over $M_{\text{inv}}$. However, the fact that $M_{\text{inv}}$  peaks around the maximum allowed in the Dalitz plot \cite{weisemi}, as we show in Fig. \ref{fig:ratiomass} for the present case, allows us to consider $C'$ constant over the whole range of $M_{\text{inv}}$.

The magnitudes $\tilde{E}_{\Xi_b}$ and $\tilde{\vec{p}}_{\Xi_b}$ in Eq. \eqref{Eq.:ratiomass} are the energies of $\Xi_b$ and its momentum in the rest frame of the $\bar\nu_l l$ pair which are given by \cite{navarra}
\begin{subequations}
\begin{align}
& \tilde{E}_{\Xi_b} = \frac{M^2_{\Xi_b}+M^2_{\text{inv}}-M^2_{\Xi_c^*}}{2 M_{\text{inv}}}, \\
& \tilde{p}_{\Xi_b} = \frac{\lambda^{\frac{1}{2}}\left(M^2_{\Xi_b},M^2_{\text{inv}},M^2_{\Xi_c^*} \right)}{2 M_{\text{inv}}},
\end{align}
\end{subequations}
with $\lambda(x,y,z)$ the ordinary K{\"a}llen function.\\

An approximate value for the ratio of the semileptonic production for the two resonances is given by
\begin{equation}
\label{Eq.:ratiosemi}
R= \frac{\Gamma_{\Xi_b \rightarrow \bar{\nu}_l l \ \Xi_c (2790)}}{\Gamma_{\Xi_b \rightarrow \bar{\nu}_l l \ \Xi_c (2815)}} = \frac{A_{\frac{1}{2}}V_{\text{had}}\left(\frac{1}{2}\right)}{A_{\frac{3}{2}}V_{\text{had}}\left(\frac{3}{2}\right)}.
\end{equation}

\section{Results}
\begin{table*}[tbp!]\centering
\captionsetup{justification=raggedright}
\caption{Values of $g$ and $gG$ for the different channels for the resonance $\Xi_c^0 (2790) (\frac{1}{2}^-)$.\label{tab:table2}}
\centering
\begin{tabular}{l|cccc}
\hline\hline
& $\Sigma \ D$  & $\Sigma \ D^*$ & $\Lambda \ D$ & $\Lambda \ D^*$ \\ \hline
$g$ & $-1.178-i0.101$ & $0.777+i0.285$ &  $-1.396+i0.892$  & $0.569-i0.601$ \\
$gG$ & $6.544+i0.239$ & $-3.372-i1.067$  &  $8.277-i5.921$ & $-2.45+i2.844$ \\
\hline\hline
\end{tabular}
\end{table*}
\begin{table}[tbp!]
\centering
\captionsetup{justification=raggedright}
\caption{Values of $g$ and $gG$ for the different channels for the resonance $\Xi_c^0 (2815) (\frac{3}{2}^-)$.\label{tab:table3}}
\begin{tabular}{l|cc}
\hline\hline
& $\Lambda \ D^*$  & $\Sigma \ D^*$ \\ \hline
$g$ & $2.346-i0.599$ & $0.791+i0.49$ \\
$gG$ & $-12.297+i4.213$ & $-4.148-i2.15$ \\
\hline\hline
\end{tabular}
\end{table}

We use the values of $g_{R,\Sigma D}, \ g_{R, \Sigma D^*}, \ g_{R, \Lambda D}, \ g_{R, \Lambda D^*}$ and of the $G_{\Sigma D}, \ G_{\Sigma D^*}, \ G_{\Lambda D}, \ G_{\Lambda D^*}$ from Ref. \cite{romanets} which we have redone in order to evaluate the complex couplings and the $G$ functions since only the modulus of $g_{R,i}$ were given there and the values of $G_i$ were not tabulated. We give all this information in Tables \ref{tab:table2} and \ref{tab:table3}.
The $G$ functions are taken from dimensional regularization subtracting the value of $G$ at $s= \alpha \left(M_{\text{th}}^2+m_{\text{th}}^2 \right)$ with $\alpha=0.9698$ and $M_{\text{th}}+m_{\text{th}}$ the mass of the lightest hadronic channel of all the coupled channels for a given quantum number \cite{recio}.
The couplings are obtained from the residues of the amplitudes at the complex pole positions $M'_R$,
\begin{equation}\label{eq:Tij}
  T_{ij}\sim \frac{g_i\,g_j}{\sqrt{s}-M'_R},
\end{equation}
\begin{equation}\label{eq:g1}
  g^2_1=\lim_{\sqrt{s} \to M'_R} (\sqrt{s}-M'_R )\;T_{11},~~~~~\frac{g_i}{g_1}=\lim_{\sqrt{s} \to M'_R} \frac{T_{i1}}{T_{11}}.
\end{equation}
One coupling, $g_1$ has arbitrary sign and the rest of the signs are defined with respect to that one. Since the amplitudes $T_{ij}$ are generally complex, so are the residues of the poles and the couplings.

Using the values in Tables \ref{tab:table2} and \ref{tab:table3} and Eq. \eqref{Eq.:ratiorate} we obtain
\begin{equation}\label{eq:R1}
R_1=\frac{\Gamma_{\Xi_b \rightarrow \pi^- \Xi_c^0(2790)}}{\Gamma_{\Xi_b \rightarrow \pi^- \Xi_c^0 (2815)}}= 0.384,
\end{equation}
and from Eq. \eqref{Eq.:ratiorate2}
\begin{equation}\label{eq:R2}
R_2=\frac{\Gamma_{\Xi_b \rightarrow D_s^- \Xi_c^0(2790)}}{\Gamma_{\Xi_b \rightarrow D_s^- \Xi_c^0 (2815)}}= 0.273.
\end{equation}
Similarly we can obtain from Eq.~\eqref{Eq.:ratiorate3}
\begin{equation}\label{eq:R3}
R_3=\frac{\Gamma_{\Xi_b \rightarrow D_s^- \Xi_c^0(2790)}}{\Gamma_{\Xi_b \rightarrow \pi^- \Xi_c^0(2790)}}=0.686.
\end{equation}

In order to see how sensitive these rates are to the values of the $D^* B$ couplings we reevaluate them by first setting them to zero or changing their sign. The results we obtain are shown in Table \ref{tab:tabledif}.

\begin{table}[tpb]
\centering
\captionsetup{justification=raggedright}
\caption{Values of $R_1, R_2, R_3$ obtained by both changing the sign of the $g_{R, B D^*}$ couplings or setting them to zero.\label{tab:tabledif}}
\begin{tabular}{l|lll}
\hline\hline
& ~~$R_1$  & ~~$R_2$ & ~~$R_3$ \\ \hline
$g_{R,\Sigma D^*}=0$ & ~$0.84$~~ & $0.596$~~ & $0.686$ \\
$g_{R,\Lambda D^*}=0$ & ~$0.205$~~ & $0.145$~~ & $0.686$ \\
$g_{R,\Sigma D^*}\rightarrow-g_{R,\Sigma D^*}$~ & ~$0.481$~~ & $0.341$~~ & $0.686$  \\
$g_{R,\Lambda D^*}\rightarrow-g_{R,\Lambda D^*}$~ & ~$0.071$~~ & $0.05$~~ & $0.686$ \\
\hline\hline
\end{tabular}
\end{table}

As we can see, the results shown in Table \ref{tab:tabledif} tell us the relevance of the $D^*B$ components in the production of these resonances.

As for the sector of the semileptonic decay rates corresponding to Eq. \eqref{Eq.:ratiosemi} we find that
\begin{equation}\label{eq:R}
R= \frac{\Gamma_{\Xi_b \rightarrow \bar{\nu}_l l \ \Xi_c (2790)}}{\Gamma_{\Xi_b \rightarrow \bar{\nu}_l l \ \Xi_c (2815)}}=0.191,
\end{equation}
and if we integrate Eq.~\eqref{Eq.:ratiomass} we find
\begin{equation}\label{eq:Rsc}
R=0.197.
\end{equation}
As we can see, the numbers are essentially the same.\\
Once again, if the couplings to $D^* B$ states are changed we obtain different results, shown in Table \ref{tab:tabledif2}.
\begin{table}[t]
\centering
\captionsetup{justification=raggedright}
\caption{Values of $R$ of the semileptonic decay, obtained by both changing the sign of the $g_{R, B D^*}$ couplings or setting them to zero.\label{tab:tabledif2}}
\begin{tabular}{l|c}
\hline\hline
& $R$ \\ \hline
$g_{R,\Sigma D^*}=0$ & ~$0.430$~ \\
$g_{R,\Lambda D^*}=0$ & $0.105$\\
$g_{R,\Sigma D^*}\rightarrow-g_{R,\Sigma D^*}$ &  $0.246$  \\
$g_{R,\Lambda D^*}\rightarrow-g_{R,\Lambda D^*}$ & $0.036$ \\
\hline\hline
\end{tabular}
\end{table}

\section{Relativistic Effects, Estimation of absolute rates and uncertainties}\label{sec:Relat_Effec}
The evaluation of rates presented in the previous section was based in a non relativistic approximation to the operator in Eq.~\eqref{eq:Lqwq}, given by Eq.~\eqref{Eq.:weaktrans}. This could look as a very drastic approximation since in the $\Xi^-_b \to \pi^- \Xi^0_c(2790)$ decay, the momentum of the $\Xi^0_c(2790)$ is $\sim 2223\;{\rm MeV}/c$, not much smaller than its mass. Yet, the difference between the relativistic and non relativistic energies is only $12\%$. But the effect of some neglected terms in the matrix element of Eq.~\eqref{eq:Lqwq} could be bigger. Actually this is the case, and in Ref.~\cite{weisemi} the relativistic effects were considered in the  $\Lambda_b \to \bar \nu_l \,l\, \Lambda_c(2595) (\Lambda_c(2625))$ semileptonic decays and the effect was an increase in about $30\%$ of the individual decay rates. Yet, when the ratios of rates were taken, the effects amounted to only about $1\%$. Here we will do this exercise again for the semileptonic decay and extend it to the nonleptonic case. Let us begin by this latter one.

Let us start from the full relativistic amplitude obtained from Eqs. \eqref{eq:LWpi}, \eqref{eq:Lqwq},
\begin{equation}\label{eq:t_rel}
t_{\rm rel} \propto q_\mu \bar{q}_{\text{fin}} \gamma^{\mu}(1-\gamma_5) q_{\text{in}} \equiv q^\mu Q_\mu.
\end{equation}
Considering the $b$ and $c$ quarks as free particles, for the purpose of estimating the effect of the relativistic terms, and summing and averaging over the spin third components (hence, also neglecting the separation into the $PB$ and $VB$ baryon components that we have done), we can write (see Eq. (8) of Ref.~\cite{navarra})
\begin{eqnarray}
  &&\overline{\sum}\sum |t_{\rm rel}|^2 \nonumber\\
   &=& q^\mu q^\nu \frac{ p_{b\mu} p_{c\nu} +p_{c\mu} p_{b\nu} -p_b \cdot p_c g_{\mu\nu}-i\epsilon_{\rho \mu \sigma \nu} p^\rho_b p^\sigma_c  }{m_b m_c} \nonumber\\
   &=& \frac{2(q\cdot p_b) (q\cdot p_c)-q^2 (p_b \cdot p_c)}{m_b m_c}. \label{eq:Sum_t_rel}
\end{eqnarray}
At this point we make use of the heavy-quark symmetry approximate relations
\begin{equation}\label{eq:pm_ratio}
  \frac{p^\mu_b}{m_b}\sim \frac{p^\mu_{\Xi_b}}{M_{\Xi_b}},~~~~~~\frac{p^\mu_c}{m_c}\sim \frac{p^\mu_{R}}{M_R},
\end{equation}
where $R$ stands for the $\Xi^*_c$ final baryon resonance produced. These relationships are obtained neglecting the internal relative three momenta of the quarks in the heavy baryons versus their masses, and are commonly used in heavy hadron dynamics. Then Eq.~\eqref{eq:Sum_t_rel} can be approximately written as
\begin{equation}\label{eq:Sum_t_rel2}
 \overline{\sum}\sum |t_{\rm rel}|^2 =\frac{2(q\cdot p_{\Xi_b})(q\cdot p_{\Xi^*_c})-q^2 (p_{\Xi_b} \cdot p_{\Xi^*_c})}{M_{\Xi_b}M_{\Xi^*_c}}.
\end{equation}
We can see that if we make the non relativistic reduction $p_{\Xi^*_c} \simeq (M_{\Xi^*_c}, \vec 0)$, then we get $\overline{\sum}\sum |t_{\rm rel}|^2 ={q^{0}}^2 +\vec{q\,}^2$, which is the $(|\vec q\,|^2 +\omega^2_\pi)$ factor that we find in Eq.~\eqref{Eq.:sum12} for $J=1/2$. For $J=3/2$ the factor is $2\omega^2_\pi$. There is only $0.2\%$ difference between these two magnitudes, but we can take just the first term in the numerator of Eq.~\eqref{eq:Sum_t_rel2}, $\frac{2(q\cdot p_{\Xi_b})(q\cdot p_{\Xi^*_c})}{M_{\Xi_b}M_{\Xi^*_c}}$, as the relativistic form for the case of spin $3/2$, replacing $2\omega^2_\pi$. The terms in Eq.~\eqref{eq:Sum_t_rel2} are trivially evaluated since
\begin{eqnarray}
   &&q^2=m^2_\pi, ~~~~~~~2q\cdot p_{\Xi^*_c} = M^2_{\Xi_b}-m^2_\pi -M^2_{\Xi^*_c}, \nonumber\\
  && 2q\cdot p_{\Xi_b} = M^2_{\Xi_b}+m^2_\pi -M^2_{\Xi^*_c},\nonumber\\
   &&2p_{\Xi_b} \cdot p_{\Xi^*_c}  = M^2_{\Xi_b} +M^2_{\Xi^*_c}-m^2_\pi.\nonumber
\end{eqnarray}
When we make these replacements in the individual rates we obtain the following results:
\begin{subequations}
\begin{align}
& \frac{\Gamma^{\text{(rel)}}_{\Xi^-_b \rightarrow \pi^- \Xi^0_c(2790)}}{\Gamma^{\text{(nonrel)}}_{\Xi^-_b \rightarrow \pi^- \Xi^0_c(2790)}}=2.07, \\
& \frac{\Gamma^{\text{(rel)}}_{\Xi^-_b \rightarrow \pi^- \Xi^0_c(2815)}}{\Gamma^{\text{(nonrel)}}_{\Xi^-_b \rightarrow \pi^- \Xi^0_c(2815)}}=2.05.
\end{align}
\end{subequations}

As we can see, the relativistic corrections are important and increase the individual rates in about a factor of two. Yet, since the ratios of rates is the only thing that we determine, we have now, replacing $R_1$ of Eq.~\eqref{eq:R1},
\begin{equation}
R_1^{\text{(rel)}}=\frac{\Gamma^{\text{(rel)}}_{\Xi^-_b \rightarrow \pi^- \Xi^0_c(2790)}}{\Gamma^{\text{(rel)}}_{\Xi^-_b \rightarrow \pi^- \Xi^0_c(2815)}}=0.389,
\end{equation}
while before $R_1$ was $0.384$. Hence, the change in the ratio is a mere $1\%$.
Similarly we evaluate
\begin{subequations}
\begin{align}
& \frac{\Gamma^{\text{(rel)}}_{\Xi^-_b \rightarrow D_s^- \Xi^0_c(2790)}}{\Gamma^{\text{(nonrel)}}_{\Xi^-_b \rightarrow D_s^- \Xi^0_c(2790)}}=1.64, \\
& \frac{\Gamma^{\text{(rel)}}_{\Xi^-_b \rightarrow D_s^- \Xi^0_c(2815)}}{\Gamma^{\text{(nonrel)}}_{\Xi^-_b \rightarrow D_s^- \Xi^0_c(2815)}}=1.52.
\end{align}
\end{subequations}
We can see that because of the larger mass of the $D_s^-$ with respect to the one of the pion, the $\Xi^0_c$ momentum is smaller and the relativistic effects are also smaller. Once again we look at the ratio $R_2$ of Eq.~\eqref{eq:R2} and we obtain now
\begin{equation}
R_2^{\text{(rel)}} = \frac{\Gamma^{\text{(rel)}}_{\Xi^-_b \rightarrow D_s^- \Xi^0_c(2790)}}{\Gamma^{\text{(rel)}}_{\Xi^-_b \rightarrow D_s^- \Xi^0_c(2815)}}=0.295,
\end{equation}
replacing the nonrelativistic value of $0.273$.
The effects in this ratio are of the order of $8\%$.

Finally, we look into the ratio $R_3$ of Eq.~\eqref{eq:R3} and we find now
\begin{equation}
R_3^{\text{(rel)}} = \frac{\Gamma^{\text{(rel)}}_{\Xi^-_b \rightarrow D_s^- \Xi^0_c(2790)}}{\Gamma^{\text{(rel)}}_{\Xi^-_b \rightarrow \pi^- \Xi^0_c(2790)}}=0.543,
\end{equation}
replacing the nonrelativistic value of $0.686$. In this case the change is of the order of $20\%$, because of the larger relativistic effects in the case of the $\pi^-$ emission compared to the one of $D_s^-$ emission.

In order to estimate the relativistic effects of the semileptonic decay we follow the steps of Ref.~\cite{weisemi}. We do not repeat the steps here but, using the results of section VI of Ref.~\cite{weisemi}, we replace in Eq.~\eqref{eq:LLQQ}
\begin{equation}\label{eq:replace_pp-1}
  p_\nu p_l \to \frac{(p_{\Xi_b} \cdot p_\nu)\, (p_{\Xi^*_c} \cdot p_l)}{M_{\Xi_b}\;M_{\Xi^*_c}},
\end{equation}
or, equivalently (see Eq.~(35) of Ref.~\cite{weisemi}) replacing the angle integrated value of $p_\nu p_l$
\begin{eqnarray}\label{eq:replace_pp-2}
  p_\nu p_l &\equiv & \frac{1}{M^2_{\Xi_b}}\left( \frac{M_{\rm inv}}{2}\right)^2  \left[  \tilde{E}^2_{\Xi_b} -\frac{1}{3} \vec{\tilde{p}}^{\,2}_{\Xi_b}\right]   \nonumber\\
   &\to& \frac{1}{M_{\Xi_b}\,M_{\Xi^*_c}} \left( \frac{M_{\rm inv}}{2}\right)^2 \left[  \tilde{E}_{\Xi_b} \tilde{E}_{\Xi^*_c} -\frac{1}{3} \vec{\tilde{p}}^{\,2}_{\Xi_b}\right],~
\end{eqnarray}
where $M_{\rm inv}$ is the $\bar \nu l$ invariant mass and the energies and momenta with tilde refer to the rest frame of the $\bar \nu l$, given by \cite{weisemi}
\begin{equation}\label{eq:tilde_p}
  \tilde{p}_{\Xi_b}= \tilde{p}_{\Xi^*_c}=\frac{\lambda^{1/2}(M^2_{\Xi_b},M^2_{\rm inv}, M^2_{\Xi^*_c})}{2M_{\rm inv}},
\end{equation}
and $\tilde{E}_{\Xi_b}=\sqrt{\tilde{p}^2_{\Xi_b}+M^2_{\Xi_b}}$, $\tilde{E}_{\Xi^*_c}=\sqrt{\tilde{p}^2_{\Xi_c}+M^2_{\Xi^*_c}}$.
When we make these replacements in the semileptonic decays we find the following results:
\begin{subequations}
\begin{align}
& \frac{\Gamma^{\text{(rel)}}_{\Xi^-_b \rightarrow \bar{\nu}_l l \Xi^0_c(2790)}}{\Gamma^{\text{(nonrel)}}_{\Xi^-_b \rightarrow \bar{\nu}_l l  \Xi^0_c(2790)}}=1.46, \\
& \frac{\Gamma^{\text{(rel)}}_{\Xi^-_b \rightarrow \bar{\nu}_l l  \Xi^0_c(2815)}}{\Gamma^{\text{(nonrel)}}_{\Xi^-_b \rightarrow\bar{\nu}_l l  \Xi^0_c(2815)}}=1.45,
\end{align}
\end{subequations}
and the ratio $R$ of Eqs.~\eqref{eq:R}, \eqref{eq:Rsc} becomes now
\begin{equation}
R=\frac{\Gamma^{\text{(rel)}}_{\Xi^-_b \rightarrow  \bar{\nu}_l l \Xi^0_c(2790)}}{\Gamma^{\text{(rel)}}_{\Xi^-_b \rightarrow  \bar{\nu}_l l \Xi^0_c(2815)}}=0.198,
\end{equation}
replacing the nonrelativistic value of $0.197$ of Eq.~\eqref{eq:Rsc}, less than $1\%$ change. The smaller relativistic effects in the case of the semileptonic decay can be traced back to the large invariant mass of the $\bar{\nu}_l l$ pair (see Fig.~\ref{fig:ratiomass}) with respect to the $\pi^-$ or even the $D_s^-$ mass.

\subsection{Estimation of absolute values for the rates and uncertainties}
The evaluation of the absolute values for the rates would require the knowledge of the form factor of Eq.~\eqref{Eq.:matel} for which we do not have enough information, particularly for the excited $c$ quark of $\phi_{\text{fin}}(r)$. This is the reason why we have calculated ratios where this matrix element will cancel. In order to evaluate absolute values for the decay rates, we shall construct ratios with respect to a related process for which there are experimental data. The ideal one is the decay $\Lambda_b \to \pi^- \Lambda_c(2595) (\, \Lambda_c(2625))$. In the case of the $\pi^- \Lambda_c(2595)$ the momentum of the $\Lambda_c$ is $q=2208\,{\rm MeV}/c$. This value only differs in $15 \,{\rm MeV}/c$ from the one of the $\Xi_b \to \pi^- \Xi_c(2790)$, less than 1\% difference. Thus, since the transition $b\to \pi^- c$ is
the same in both cases and the $ds$ or $ud$ quarks are spectators in the $\Xi_b$  and $\Lambda_b$ decays respectively, we can simply assume the matrix ${\rm ME}(q)$
of Eq.~\eqref{Eq.:matel} to be the same in both reactions. In that case, we have
\begin{equation}\label{eq:BR_ratio}
  \frac{BR(\Xi_b \to \pi^-\, \Xi^*_c)}{BR(\Lambda_b \to \pi^-\, \Lambda^*_c)}
  = \frac{M_{\Xi^*_c}}{M_{\Xi_b}}\; \frac{M_{\Lambda_b}}{M_{\Lambda^*_c}} \;\frac{q \;\overline{\sum}\sum |t|^2 \Big|_{\Xi_{b}}}{q \;\overline{\sum}\sum |t|^2 \Big|_{\Lambda_b}}\cdot \frac{\Gamma_{\Lambda_b}}{\Gamma_{\Xi_b}},
\end{equation}
where $\overline{\sum}\sum |t|^2 \Big|_{\Xi_{b}}$ is given by Eqs.~\eqref{Eq.:sum12}, \eqref{Eq.:sum32} and $\overline{\sum}\sum |t|^2 \Big|_{\Lambda_b}$ by Eqs.~(41), (42) of Ref.~\cite{weihong} which we write below
\begin{eqnarray}\label{eq:t2_Lambdab}
  &J=\frac{1}{2}:& ~ \overline{\sum}\sum |t|^2 \Big|_{\Lambda_b} \!= (q^2+w^2_\pi) \left|
  \frac{1}{2} g_{R, DN}\, G_{DN} \right.\nonumber\\
  &&~~~~~~~~~~~~~~~~~~~\left.+\frac{1}{2\sqrt{3}} g_{R,D^*N}\,G_{D^* N} \right|^2, \\
  &J=\frac{3}{2}:& ~ \overline{\sum}\sum |t|^2 \Big|_{\Lambda_b} \!=\! 2w^2_\pi \left|
  \frac{1}{\sqrt{3}}  G_{D^* N}\, g_{R,D^*N} \right|^2.~~~~~
\end{eqnarray}

In Eq.~\eqref{eq:BR_ratio}, we could take~\cite{pdg},
\begin{equation}\label{eq:ratio_Gamma}
  \frac{\Gamma_{\Lambda_b}}{\Gamma_{\Xi_b}}=\frac{\tau_{\Xi_b}}{\tau_{\Lambda_b}}=1.08\pm 0.19.
\end{equation}
Using the following ratios \cite{pdg}
\begin{eqnarray}
  \!\!BR[\Lambda_b \to \pi^- \Lambda_c(2595)]\!\! &=&\!\! \frac{(3.4\pm 1.5 )\times 10^{-4}}{BR[\Lambda_c(2595) \to \Lambda_c \pi^+ \pi^-]},~~~~ \label{eq:BR_Lambdab1}\\
   \!\!BR[\Lambda_b \to \pi^- \Lambda_c(2625)]\!\! &=&\!\! \frac{(3.3\pm 1.3 )\times 10^{-4}}{BR[\Lambda_c(2625) \to \Lambda_c \pi^+ \pi^-]},\label{eq:BR_Lambdab2}~~~~
\end{eqnarray}
with $BR[\Lambda^*_c \to \Lambda_c \pi^+ \pi^-]=0.67$ \cite{pdg}, we obtain
\begin{eqnarray}\label{eq:BR_Xib}
  BR[\Xi_b \to \pi^- \Xi_c(2790)] &=& (7\pm 4)\times 10^{-6},  \label{eq:BR_Xia1}\\
  BR[\Xi_b \to \pi^- \Xi_c(2815)] &=& (13\pm 7)\times 10^{-6},  \label{eq:BR_Xia2}
\end{eqnarray}
where the 50\% relative error is obtained summing in quadratures the relative errors in Eqs.~\eqref{eq:ratio_Gamma},  \eqref{eq:BR_Lambdab1}, \eqref{eq:BR_Lambdab2} and an error of the order of 20\% affecting to
the $\Lambda_b \to \pi^-\, \Lambda^*_c$ decay, as discussed in Ref.~\cite{weihong}. It estimates the effects produced by the $D_s \Lambda$ and  $D^*_s \Lambda$ channels neglected in the approach followed in that work (see discussion in Section 6 of that reference).

As for the semileptonic decay, we would equally have
\begin{equation}\label{eq:BR_XibtoLamb_semilep}
  \frac{BR(\Xi_b \to \bar \nu_l \,l\,\Xi^*_c)}{BR(\Lambda_b \to \bar \nu_l \, l\, \Lambda^*_c)}=\frac{\int {\rm d} M_{\rm inv} \; \frac{{\rm d}\Gamma}{{\rm d}M_{\rm inv}} \Big|_{\Xi_b}}{\int {\rm d} M_{\rm inv}\; \frac{{\rm d}\Gamma}{{\rm d}M_{\rm inv}}\Big|_{\Lambda_b}} \cdot \frac{\Gamma_{\Lambda_b}}{\Gamma_{\Xi_b}},
\end{equation}
where $\frac{{\rm d}\Gamma}{{\rm d}M_{\rm inv}}\Big|_{\Xi_b}$ is given by Eq.~\eqref{Eq.:ratiomass} and $\frac{{\rm d}\Gamma}{{\rm d}M_{\rm inv}}\Big|_{\Lambda_b}$ by Eq.~(27) of Ref.~\cite{weisemi}, which we reproduce below, with
\begin{equation}\label{eq:dGamma_Lamb}
 \frac{{\rm d}\Gamma}{{\rm d}M_{\rm inv}}\Big|_{\Lambda_b}=\frac{M_{\Lambda^*_c}}{M_{\Lambda_b}}\;2m_\nu \;2m_l \frac{1}{(2\pi)^3}\; p_{\Lambda^*_c} \; \tilde{p}_l \; \overline{\sum}\sum |T|^2 \Big|_{\Lambda_b}
\end{equation}
with $p_{\Lambda^*_c}$ the $\Lambda^*_c$ momentum in the $\Lambda_b$ rest frame and $\tilde{p}_l$ the lepton momentum in the $\bar \nu l$ rest frame, and
\begin{equation}\label{eq:T2_Lamb}
 \overline{\sum}\sum |T|^2 \Big|_{\Lambda_b}=C'^2 \frac{8}{m_\nu m_l}\, p_\nu\, p_l \; A_JV_{\rm had} (J)
\end{equation}
with
\begin{align}\label{eq:AJVhad}
   & A_J V_{\rm had} (J) \nonumber\\
  \equiv & \left\{
  \begin{array}{ll}
    \left|  \frac{1}{2}  g_{R,DN}\, G_{DN}  + \frac{1}{2\sqrt{3}} g_{R,D^*N}\,G_{D^*N}  \right|^2, & {\rm for~} J=1/2 \\[3mm]
    2\left| \frac{1}{\sqrt{3}} g_{R,D^*N}\, G_{D^*N}  \right|^2, & {\rm for~} J=3/2\\
    \end{array}
   \right.
\end{align}
The experimental branching ratios are \cite{pdg}
\begin{eqnarray}
  BR[\Lambda_b \to \bar \nu_l l \Lambda_c(2595)] &=& \left( 7.9 ^{+4.0}_{-3.5} \right) \times 10^{-3}, \label{eq:BR_Lamb1_PDG} \\
  BR[\Lambda_b \to \bar \nu_l l \Lambda_c(2625)] &=& \left( 13.0 ^{+6.0}_{-5.0} \right) \times 10^{-3},\label{eq:BR_Lamb2_PDG}
\end{eqnarray}
from where we obtain
\begin{eqnarray}\label{eq:BR_PDG}
  BR[\Xi_b \to \bar \nu_l l \Xi_c(2790)] &=& \left( 1.0  ^{+0.6}_{-0.5}  \right) \times 10^{-4}, \label{eq:BR_Xib1} \\
  BR[\Xi_b \to \bar \nu_l l \Xi_c(2815)] &=& \left( 3.3 ^{+1.8}_{-1.6} \right) \times 10^{-4},\label{eq:BR_Xib2}
\end{eqnarray}
where the 50-60\% relative error comes from summing in quadratures the relative errors of Eq.~\eqref{eq:ratio_Gamma}, Eqs.~\eqref{eq:BR_Lamb1_PDG}, \eqref{eq:BR_Lamb2_PDG} and an extra 20\% from the consideration of the $D_s \Lambda$, $D^*_s \Lambda$ channels in Ref.~\cite{weisemi}.

We have also estimated uncertainties in the magnitudes that we have calculated, related to uncertainties in the model. For this, we have used the freedom that we have in the cut off, or subtraction constant in dimensional regularization, employed to regularize the loops. We have allowed small changes that induce a change of about 6 MeV in the mass of the ${\Xi_c^0}^*$ states (about double than the empirical errors). With this we find the uncertainties:
\begin{subequations}
\begin{align}
& \frac{\delta R_1}{R_1} \simeq 0.35, \\
& \frac{\delta R_2}{R_2} \simeq 0.35, \\
& \frac{\delta R_3}{R_3} \simeq 0, \\
& \frac{\delta R}{R} \simeq 0.35 .
\end{align}
\end{subequations}

As to the absolute values in Eqs.~\eqref{eq:BR_Xia1} \eqref{eq:BR_Xia2} \eqref{eq:BR_Xib1} \eqref{eq:BR_Xib2} we find uncertainties also of the order of 25\% from this source, which summed in quadratures to the existing errors, do not change much the errors that we already associated to these numbers and discussed above. It might be surprising that the errors in the ratios are bigger than in the absolute values of the rates from this source. This is because an increase in the subtraction constant decreases the rate for the $\Xi_c(2790)$ and increases the rate for the $\Xi_c(2815)$ both in the nonleptonic and the semileptonic decays.

We want to note that the smaller absolute numbers obtained for the present decay, compared to those of the $\Lambda_b$  stem from the large cancellations between the terms in Eqs.~\eqref{Eq.:sum12} \eqref{Eq.:sum32} and  \eqref{Eq.:sum122} \eqref{Eq.:sum322}, between the $\Sigma D$ and $\Lambda D$ contributions. We should also warn that to estimate the absolute rates we have used two different theoretical models for the $DN$, $D^*N$ and $D\Sigma$, $D^* \Sigma$, $D \Lambda$, $D^* \Lambda$ interactions from Ref.~\cite{uchinohid} and Ref.~\cite{romanets} respectively. One should expect some systematic errors from this source, more difficult to evaluate, but we think that, with the large uncertainties that we already have, these new uncertainties would also be accommodated.

\section{Conclusion}

We have studied the nonleptonic $\Xi_b^- \rightarrow M + \Xi_c^*$, with $M= \pi^-, \ D_s^-$ and $\Xi_c^* = \Xi_c^0 (2790) ( \frac{1}{2}^- )$,  $\Xi_c^0 (2815) ( \frac{3}{2}^- )$. We have assumed that the $\Xi_c^*$ resonances are dynamically generated from the $PB$ and $VB$ interactions, as done in Ref. \cite{romanets}. We saw that the present decays only involved the $D \Lambda, \ D \Sigma, \ D^* \Lambda, \ D^* \Sigma$ channels and we took the needed couplings from that work. Given the fact that the momentum of the meson $M$ is very similar for the case of the production of the two resonances (since their masses are very close) we could eliminate in the ratio of widths the matrix element at the quark level involving the wave functions of the $b$ and $c$ quarks. Then, only factors related to the spin structure of the channels and the couplings of the hadronic model for the resonances were relevant, which tells us that the measurement of these partial decay widths are relevant to learn details on the nature of
the $\Xi_c^*$ resonances. With more uncertainty we were able to also predict the ratio of $\Xi_b^- \rightarrow \pi^- \Xi_c^*$ and $\Xi_b^- \rightarrow D_s^- \Xi_c^*$ for the same resonance.

We also evaluated the semileptonic rates. In this case we can only evaluate one ratio, the one of the semileptonic decay $\Xi_b \rightarrow \bar{\nu}_l l \Xi_c^*$ for the $\Xi_c^0 (2790)$ and $\Xi_c^0 (2815)$ resonances. Once again, the predictions will be valuable when these partial decay widths can be measured. We should stress that both the nonleptonic and semileptonic decay widths are measured for the case of $\Lambda_b \rightarrow \pi^- \Lambda_c (2595), \ \Lambda_b \rightarrow \pi^- \Lambda_c (2625)$ and $\Lambda_b \rightarrow \bar{\nu}_l l \Lambda_c (2595)$ and $\Lambda_b \rightarrow \bar{\nu}_l l \Lambda_c (2625)$ and the method used here gave results in agreement with experiment \cite{weihong,weisemi}, so we are confident that the predictions done here are fair.
We also estimated the absolute branching ratios of all these decays from the ratios to the related $\Lambda_b \to \pi^- \Lambda_c(2595)(\Lambda_c(2625))$, $\Lambda_b \to \bar \nu_l \,l \Lambda_c(2595)(\Lambda_c(2625))$ reactions and the experimental rates for these latter decays. The branching ratios obtained are well within measurable range, where branching ratios of $\Xi_b^-$ of the order of $10^{-7}$ have already been observed \cite{pdg}.
In any case the experimental result could test the accuracy of the model of Ref. \cite{romanets}, which is one of the possible ways to address the molecular states, with a particular dynamics consistent with HQSS.

We also checked that the results were sensitive to the couplings of the $D^* B$ components and confirmation of this feature by experiment could give a boost to the relevance of the mixing of pseudoscalar-baryon and vector-baryon components in the building up of the molecular baryonic states, a subject which is catching up in the hadronic community \cite{javier,romanets,juanmas,kemchan1,kemchan2,uchinoop,uchinohid}.

\section*{Acknowledgments}
R. P. Pavao wishes to thank the Generalitat Valenciana in the program Santiago Grisolia.
This work is partly supported by the
National Natural Science Foundation of China under Grants No. 11565007, No. 11647309 and No. 11547307.
This work is also partly supported by the Spanish Ministerio
de Economia y Competitividad and European FEDER funds
under the contract numbers  FIS2014-51948-C2-1-P, and FIS2014-51948-C2-2-P, and the Generalitat Valenciana
in the program Prometeo II-2014/068.

\bibliographystyle{plain}

\begin{thebibliography}{999}

\bibitem{ecker}
G. Ecker. {\it Progress in Particle and Nuclear Physics}, 35:1-80, 1995.

\bibitem{ulfg}
V.~Bernard, N.~Kaiser and Ulf-G.~Mei{\ss}ner,
  Int.\ J.\ Mod.\ Phys.\ E {\bf 4}, 193 (1995)
  [hep-ph/9501384].

\bibitem{weise}
 N.~Kaiser, P.~B.~Siegel and W.~Weise,
  Nucl.\ Phys.\ A {\bf 594}, 325 (1995)
  [nucl-th/9505043].

\bibitem{angels}
 E.~Oset and A.~Ramos,
  Nucl.\ Phys.\ A {\bf 635}, 99 (1998)
  [nucl-th/9711022].


\bibitem{ollerulf}
 J.~A.~Oller and Ulf-G.~Mei{\ss}ner,
  Phys.\ Lett.\ B {\bf 500}, 263 (2001)
  [hep-ph/0011146].

\bibitem{carmen}
 C.~Garc\'{\i}a-Recio, J.~Nieves, E.~Ruiz Arriola and M.~J.~Vicente Vacas,
  Phys.\ Rev.\ D {\bf 67}, 076009 (2003)
  [hep-ph/0210311].

\bibitem{hyodo}
 T.~Hyodo and D.~Jido,
  Prog.\ Part.\ Nucl.\ Phys.\  {\bf 67}, 55 (2012)
  [arXiv:1104.4474 [nucl-th]].

\bibitem{jido}
D.~Jido, J.~A.~Oller, E.~Oset, A.~Ramos and U.~G.~Meissner,
  Nucl.\ Phys.\ A {\bf 725}, 181 (2003)
  [nucl-th/0303062].

\bibitem{ulfg2}
Ulf-G.~Mei{\ss}ner and T. Hyodo, {\it Pole structure of the $\Lambda(1405)$ region}, in
C. Patrignani et al. [Particle Data Group] Chin. Phys. C, {\bf 40}, 100001 (2016).

\bibitem{angelsvec}
E.~Oset and A.~Ramos,
  Eur.\ Phys.\ J.\ A {\bf 44}, 445 (2010)
  [arXiv:0905.0973 [hep-ph]].

\bibitem{sarkarvec}
S.~Sarkar, B.~X.~Sun, E.~Oset and M.~J.~Vicente Vacas,
  Eur.\ Phys.\ J.\ A {\bf 44}, 431 (2010)
  [arXiv:0902.3150 [hep-ph]].


\bibitem{hidden1}
M.~Bando, T.~Kugo and K.~Yamawaki,
  Phys.\ Rept.\  {\bf 164}, 217 (1988).

\bibitem{hidden2}
M.~Harada and K.~Yamawaki,
  Phys.\ Rept.\  {\bf 381}, 1 (2003)
  [hep-ph/0302103].

\bibitem{hidden4}
Ulf-G.~Mei{\ss}ner,
  Phys.\ Rept.\  {\bf 161}, 213 (1988).

\bibitem{javier}
E.~J.~Garzon and E.~Oset,
  Eur.\ Phys.\ J.\ A {\bf 48}, 5 (2012)
  [arXiv:1201.3756 [hep-ph]].

\bibitem{uchinoop}
W.~H.~Liang, T.~Uchino, C.~W.~Xiao and E.~Oset,
  Eur.\ Phys.\ J.\ A {\bf 51}, no. 2, 16 (2015)
  [arXiv:1402.5293 [hep-ph]].

\bibitem{uchinohid}
T.~Uchino, W.~H.~Liang and E.~Oset,
  Eur.\ Phys.\ J.\ A {\bf 52}, no. 3, 43 (2016)
  [arXiv:1504.05726 [hep-ph]].

\bibitem{romanets}
O.~Romanets, L.~Tolos, C.~Garcia-Recio, J.~Nieves, L.~L.~Salcedo and R.~G.~E.~Timmermans,
  Phys.\ Rev.\ D {\bf 85}, 114032 (2012)
  [arXiv:1202.2239 [hep-ph]].

\bibitem{lutz}
J.~Hofmann and M.~F.~M.~Lutz,
  Nucl.\ Phys.\ A {\bf 763}, 90 (2005)
  [hep-ph/0507071].

\bibitem{mizutani}
T.~Mizutani and A.~Ramos,
  Phys.\ Rev.\ C {\bf 74}, 065201 (2006)
  [hep-ph/0607257].

\bibitem{weihong}
W.~H.~Liang, M.~Bayar and E.~Oset,
  Eur.\ Phys.\ J.\ C {\bf 77}, no. 1, 39 (2017)
  [arXiv:1610.08296 [hep-ph]].

\bibitem{weisemi}
 W.~H.~Liang, E.~Oset and Z.~S.~Xie,
  Phys.\ Rev.\ D {\bf 95}, no. 1, 014015 (2017)
  [arXiv:1611.07334 [hep-ph]].

\bibitem{liang}
W.~H.~Liang and E.~Oset,
  Phys.\ Lett.\ B {\bf 737}, 70 (2014)
  [arXiv:1406.7228 [hep-ph]].

\bibitem{mai}
L.~Roca, M.~Mai, E.~Oset and Ulf-G.~Mei{\ss}ner,
  Eur.\ Phys.\ J.\ C {\bf 75}, no. 5, 218 (2015)
  [arXiv:1503.02936 [hep-ph]].

\bibitem{lhcb}
 R.~Aaij {\it et al.} [LHCb Collaboration],
  Phys.\ Rev.\ Lett.\  {\bf 115}, 072001 (2015)
  [arXiv:1507.03414 [hep-ex]].

\bibitem{miyahara}
K.~Miyahara, T.~Hyodo and E.~Oset,
  Phys.\ Rev.\ C {\bf 92}, no. 5, 055204 (2015)
  [arXiv:1508.04882 [nucl-th]].

\bibitem{feijoo}
A.~Feijoo, V.~K.~Magas, A.~Ramos and E.~Oset,
  Phys.\ Rev.\ D {\bf 92}, no. 7, 076015 (2015)
  [arXiv:1507.04640 [hep-ph]].

\bibitem{close}
F. E. Close. {\it An Introduction to Quarks and Partons}.
Academic press, 1979.

\bibitem{Miyahara:2016yyh}
  K.~Miyahara, T.~Hyodo, M.~Oka, J.~Nieves and E.~Oset,
   Phys.\ Rev.\ C {\bf 95}, 035212 (2017)
   [arXiv:1609.00895 [nucl-th]].

\bibitem{gasser}
J.~Gasser and H.~Leutwyler,
  Nucl.\ Phys.\ B {\bf 250}, 465 (1985).

\bibitem{scherer}
Stefan Scherer.
{\it Introduction to chiral perturbation theory.}
  Adv.\ Nucl.\ Phys.\  {\bf 27}, 277 (2003)
  [hep-ph/0210398].

\bibitem{oliver}
A.~Le Yaouanc, L.~Oliver, O.~Pene and J.~C.~Raynal,
  Phys.\ Rev.\ D {\bf 8}, 2223 (1973).

\bibitem{brown}
G. E. Brown. {\it Unified Theory of Nuclear Models and
Forces}. North-Holand Publishing Company, 1971.

\bibitem{preston}
M. A. Preston and R. K. Bhaduri. {\it Structure of the Nu-
cleus}. Addison Wesley Publishing Company, 1975.

\bibitem{rose}
M. E. Rose. {\it Elementary theory of angular momentum}.
Dover publications, 1995.

\bibitem{navarra}
 F.~S.~Navarra, M.~Nielsen, E.~Oset and T.~Sekihara,
  Phys.\ Rev.\ D {\bf 92}, no. 1, 014031 (2015)
  [arXiv:1501.03422 [hep-ph]].

\bibitem{sekihara}
T.~Sekihara and E.~Oset,
  Phys.\ Rev.\ D {\bf 92}, no. 5, 054038 (2015)
  [arXiv:1507.02026 [hep-ph]].

\bibitem{ikeno}
N.~Ikeno and E.~Oset,
  Phys.\ Rev.\ D {\bf 93}, no. 1, 014021 (2016)
  [arXiv:1510.02406 [hep-ph]].

\bibitem{mandl}
F.~Mandl and G.~Shae,
  {\it Quantum Field Theory.}
  John Wiley and sons, 1984



\bibitem{recio}
 C.~Garcia-Recio, V.~K.~Magas, T.~Mizutani, J.~Nieves, A.~Ramos, L.~L.~Salcedo and L.~Tolos,
  Phys.\ Rev.\ D {\bf 79}, 054004 (2009)
  [arXiv:0807.2969 [hep-ph]].

\bibitem{pdg}
C.~Patrignani {\it et al.} [Particle Data Group],
  Chin.\ Phys.\ C {\bf 40}, no. 10, 100001 (2016).

\bibitem{juanmas}
 C.~Garcia-Recio, J.~Nieves and L.~L.~Salcedo,
  Phys.\ Rev.\ D {\bf 74}, 034025 (2006)
  [hep-ph/0505233].

\bibitem{kemchan1}
 K.~P.~Khemchandani, A.~Martinez Torres, H.~Nagahiro and A.~Hosaka,
  Nucl.\ Phys.\ A {\bf 914}, 300 (2013).

\bibitem{kemchan2}
K.~P.~Khemchandani, A.~Martinez Torres, F.~S.~Navarra, M.~Nielsen and L.~Tolos,
  Phys.\ Rev.\ D {\bf 91}, 094008 (2015)
  [arXiv:1406.7203 [nucl-th]].



\end{thebibliography}

\end{document}